\documentclass{sig-alternate-05-2015}
\usepackage[utf8]{inputenc}
\usepackage{multicol}
\usepackage{amsmath}
\usepackage{mathtools}
\usepackage{graphicx}
\usepackage{float}
\usepackage{adjustbox}
\usepackage{enumitem}
\usepackage{multirow}
\usepackage{balance}
\usepackage{lastpage}

\begin{document}

\CopyrightYear{2016}
\setcopyright{rightsretained}
\conferenceinfo{HT '16}{July 10-13, 2016, Halifax, NS, Canada}
\isbn{978-1-4503-4247-6/16/07}
\doi{http://dx.doi.org/10.1145/2914586.2914611}

\title{E\^3 : Keyphrase based News Event Exploration Engine}

\numberofauthors{1}
\author{
\alignauthor
Nikita Jain 		~~   Swati Gupta 		~~   Dhaval Patel \\
\affaddr{Department of Computer Science and Enginnering}\\
\affaddr{Indian Institute of Technology Roorkee, India}\\
\email{nk27jain@gmail.com, \{sg123.dcs2014, patelfec\}@iitr.ac.in}
}

\maketitle

\begin{abstract}
This paper presents a novel system E$^3$ for extracting keyphrases from news content for the purpose of offering the news audience a broad overview of news events, with especially high content volume. Given an input query, E$^3$ extracts keyphrases and enrich them by tagging, ranking and finding role for frequently associated keyphrases. Also, E$^3$ finds the novelty and activeness of keyphrases using news publication date, to identify the most interesting and informative keyphrases.
\end{abstract}

\section{Introduction}
News media are publishing ideas, events and opinions in an increasingly wide range of data formats such as news articles, headlines, videos, tweets, hashtags and others. The explosion of Big news data has sparked the text and data mining research communities to focus on developing systems for news data exploration and analysis. Broadly, two types of news data exploration systems are developed till date: \textit{Event centric} (GDELT \cite{D:8}, EventRegistry \cite{D:7}) and \textit{Content centric} (STICS \cite{D:9}, EMM \cite{D:10}). In Event centric system, input query maps to real world events, whereas, the content centric system outputs related news articles of a given query.       

Although, both types of systems provide up-to-date news information in real time, but they overload the user with the large amounts of results. For instance, given input query ``2014 FIFA World Cup'', event centric EventRegistry suggested 11,504 news events, and the content centric STICS suggested 1,286,369 news articles having multiple organizations, people and places mentioned. Clearly, there is a need of a system that enables readers to get a broad overview of the news data generated in response of user query.  

In this paper, we propose a keyphrase based news exploration engine E$^3$ to summarize high volume news data. In our context, keyphrase is a short and meaningful chunk of text that describes an important news concepts, news entities, etc. For instance, ``Bihar election'', ``Bihar bjp'' are examples of keyphrases. Our proposed work is keyphrase centric as recent literature has shown that Keyphrase mining is able to generate a wide range of informational and important phrases from large documents (KEA \cite{D:1}, Micro-ngram \cite{D:2}, ToPMine \cite{D:3}, and SegPhrase \cite{D:4}). As shown in Figure~\ref{fig:architecture}, engine E$^3$ works in two phases: keyphrase extraction followed by keyphrase enrichment. Keyphrase extracation is performed on multi-form news (like article, keywords, and others). As existing keyphrase mining approaches performs poorly on news data, we propose a novel keyphrase extraction technique that leverages linguistic-syntactic feature, and performs very well on short texts like news headlines and video captions. The keyphrase enrichment phase finds important and interesting information related to the keyphrase, such as connected entities, novel (emerging), and active news concepts, and the role played by most frequent entities present in set of keyphrases.

\begin{figure*}[ht!]
\centering
\includegraphics[scale=0.5]{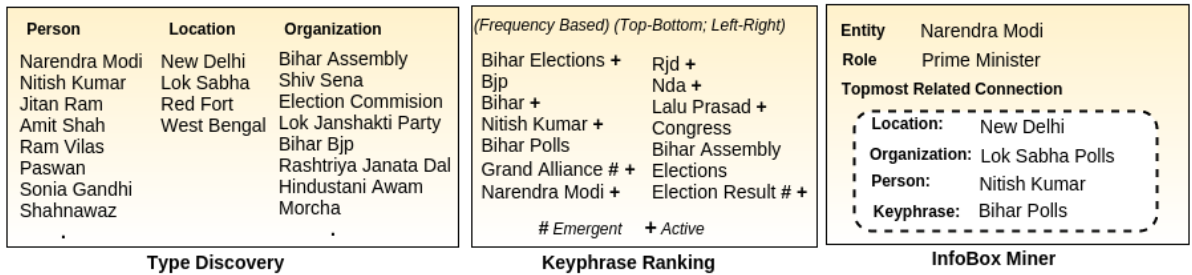}
\vspace{-0.1in}
\caption{$E^3$ System Working Example for $q$: ``Bihar Election''}
\vspace{-0.2in}
\label{fig:example}
\end{figure*}

In summary, given an input query $q$, engine E$^3$ generates a keyphrase template, as shown in Figure~\ref{fig:example}, where keyphrases are organized into three sections: Type Discovery, Keyphrase Ranking and InfoBox Miner. The keyphrase of type Person, Location and Organization are kept in Type section. The Novel, Active and Frequent keyphrases are annotated in Keyphrase Ranking section. Information like Role played, top most associated Types and Keyphrase for a selected keyphrase is stored in InfoBox section.

\section{SYSTEM OVERVIEW}

\begin{figure}[!ht]
\centering
\includegraphics[scale=0.17]{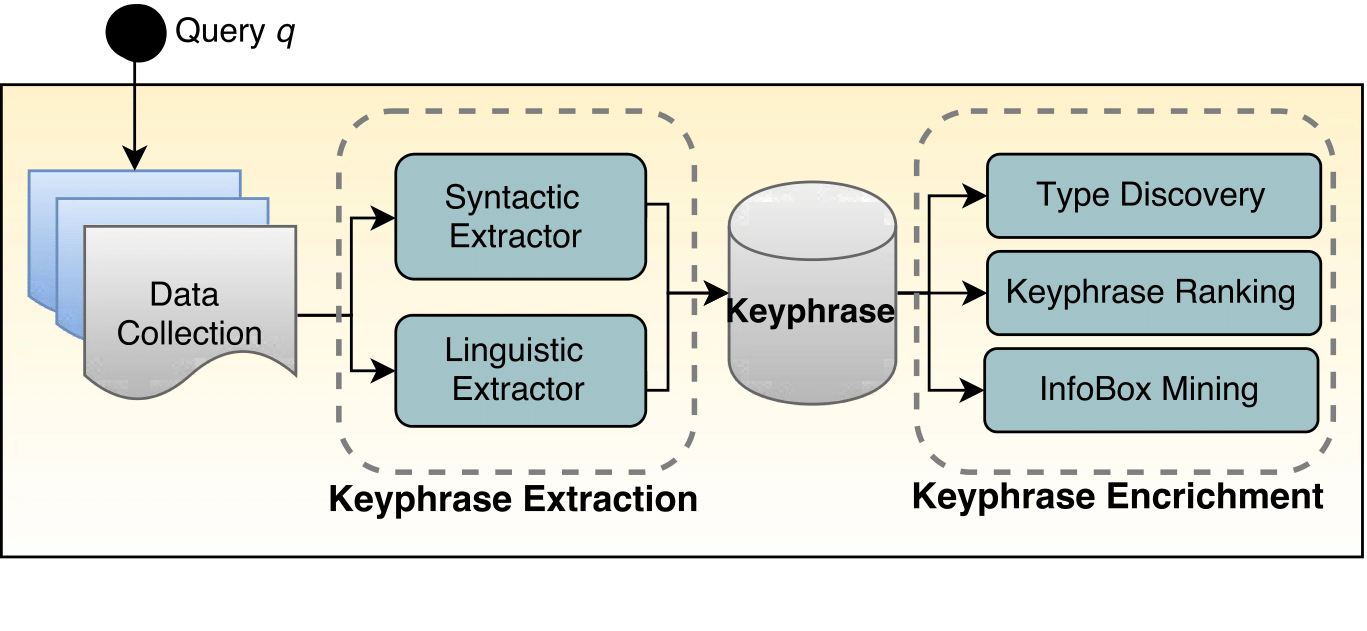}
\vspace{-0.2in}
\caption{$E^3$ System Architecture}
\vspace{-0.2in}
\label{fig:architecture}
\end{figure}

Figure~\ref{fig:architecture} gives an overview of E$^3$ architecture. 

\subsection{Data Collection}
For input query $q$, our data collection module prepares related news data published by news media. The module can either use news search engines like Google News, Yahoo News or online news structured repositories like GDELT \cite{D:8}, EMM \cite{D:10}, iMM \cite{D:6}. We use our in-house iMM system which periodically extracts news headlines (video title) along with their URL, publication date and meta-keywords. To prepare news data related to $q$, we select URL, if URL's headlines (video title) or URL's meta-keywords contains $q$.

In summary, for given query $q$, we prepare a dataset $R$ containing several news records $R_q$, where each record is described by $Quintuple$ \{Headline, Keywords, Meta-description, Article, Publication Date\}. For ``Bihar Election'' query, we retrieved 216 records from iMM for further processing. 

\subsection{Keyphrase Extraction}
Next, we extract keyphrases from all the records of news data $R$. As meta-keywords are small group of meaningful words, a naive solution is to output the meta-keywords as keyphrases. However, not all records in $R$ contain meta-keyword. As a result, meta-keywords are not sufficient enough to describe the news data completely. For example, around 20\% news headlines, obtained for ``Bihar Election'', do not have a meta-keyword. Hence, we require an efficient selection of the relevant phrases.

We observed that news headlines are short in length and contains special tokens such as colon, apostrophe, quotes, hash, dash to emphasize important information. On the other hand, meta-descriptions and news articles are long passage texts and are governed by grammatical rules. Thus, we propose two different keyphrase extractors to handle both kinds of writing styles.
\begin{itemize}[leftmargin=*]
\item \texttt{Syntactic Extractor} utilises special characters such as colon (:), apostrophe ($'$), quotes (", $'$), hash (\#), dash (-) for keyphrase extraction. In case of colon (dash), the news headline is tokenized into two parts using colon (dash) and both parts are declared as keyphrases. In case of quotes (hash) the part of text enclosed inside the quotes and hashtag containing the hash is declared as keyphrase respectively.

\item \texttt{Linguistic Extractor} applies language specific part of speech (POS) tagging on meta-description and article and then annotate collocated nouns, adjectives, noun apostrophe ($'$) connector and numbers present in the input text. These annotated tokens are further used as keyphrases.

\end{itemize}

At the end, when all the news records in $R$ are processed, we obtain a set of keyphrases $R_k$, along with the number of times they are generated. For ``Bihar Election'' query, we obtained around 6000 keyphrases for further processing. 

\subsection{Keyphrase Enrichment}
The size of generated keyphrases $R_k$ may be large and noisy. To resolve this problem, our Keyphrase Enrichment module helps in extracting valuable and actionable information by filtering and ranking the extracted keyphrases. The keyphrases are filtered using news media specific stopwords such as update, video, photo, pti and others. Next, we apply case normalization and remove duplicate keyphrases. At this point noisy keyphrases are removed. The remaining keyphrases are passed through the Type discovery, Keyphrase ranking and InfoBox mining modules.

\begin{itemize}[leftmargin=*]
\item In \texttt{Type\ Discovery} module, NER tagger is used to classify keyphrase into three types: $Person$, $Location$ and $Organization$. As existing NER taggers do not perform well on Indian named entities, we use a separate list\footnote{https://github.com/NikkiJain09/Transliteration} for Indian named entities, prepared through in-house research work. A keyphrase without any above NER type, are termed as a $News\ Concept$. For ``Bihar Election'' query, a sample keyphrases for each type is shown in Figure~\ref{fig:example}.       

\item Using \texttt{Keyphrase\ Ranking} module, keyphrases are organized according to the value of frequency, novelty and activeness. The frequency of the keyphrase is already computed during keyphrase extraction process. To compute the value of novelty and activeness, we first extracts the time intervals $q_t$ of an input query $q$, during which the $q$ was highly popular in news headlines. Next, a keyphrase is \textit{novel} (denoted by \# in Figure~\ref{fig:example}) if its frequency is very high in news headlines only during $q_t$. Similarly, a keyphrase is \textit{active} (denoted by + in Figure~\ref{fig:example}) if its frequency is very high around $q_t$. For instance, ``Grand Alliance'' and ``NDA'' are discovered as novel and active keyphrases respectively for ``Bihar Election'' query.     

\item \texttt{InfoBox\ Miner} discovers personalized information for selected keyphrase. The InfoBox displays role of keyphrase $k$ that it played with respect to query $q$. The type-wise top most connections $k$ have, determined with help of co-occurrence value of the keyphrases in $R_k$. A phrase frequently located near $k$ in the collected news corpus is labeled as the $k$'s role. Generally, keyphrases with type person and organization are preferred for InfoBox mining. Figure~\ref{fig:example} shows InfoBox for entity ``Narendra Modi'' for query ``Bihar Election''.

\end{itemize}

\section{Summary}
The engine is tested for varying the input query ranging from general topics (e.g., Election, ISIS) to specific topics (e.g., Paris Attack, Gravitational Wave) and compared the results\footnote{goo.gl/yoLXTh} with KEA, ToPMine and Micro-ngram. We found that our system outperforms existing approaches in terms of quality and quantity of keyphrases generated. As our engine is online, we can demonstrate the working to the conference participants.

\bibliography{ACM_Hyper_Text-arXiv}

\begin{thebibliography}{1}

\bibitem{D:8}
K.~Leetaru and P.~A. Schrodt, {\em GDELT: Global data on events, location, and
  tone}.
\newblock International Studies Association Annual Convention, 2013.

\bibitem{D:7}
G.~Leban, B.~Fortuna, J.~Brank, and M.~Grobelnik, {\em Event Registry: Learning
  About World Events from News}.
\newblock World Wide Web, 2014.

\bibitem{D:9}
J.~Hoffart, D.~Milchevski, and G.~Weikum, {\em STICS: Searching with Strings,
  Things, and Cats}.
\newblock Special Interest Group on Information Retrieval, 2014.

\bibitem{D:10}
R.~Steinberger, B.~Pouliquen, and E.~V. der Goot, {\em An introduction to the
  Europe Media Monitor family of applications}.
\newblock Special Interest Group on Information Retrieval, 2009.

\bibitem{D:1}
O.~Medelyan and I.~H. Witten, {\em Thesaurus Based Automatic Keyphrase
  Indexing}.
\newblock Joint Conference on Digital Libraries, 2006.

\bibitem{D:2}
K.~Wang, C.~Thrasher, E.~Viegas, X.~Li, and B.-j.~P. Hsu, {\em An Overview of
  Microsoft Web N-gram Corpus and Applications}.
\newblock North American Chapter of the Association for Computational
  Linguistics: Human Language Technologies, Association for Computational
  Linguistics, 2010.

\bibitem{D:3}
A.~El{-}Kishky, Y.~Song, C.~Wang, C.~R. Voss, and J.~Han, {\em Scalable Topical
  Phrase Mining from Text Corpora}.
\newblock Very Large Data Bases, 2014.

\bibitem{D:4}
J.~Liu, J.~Shang, C.~Wang, X.~Ren, and J.~Han, {\em Mining Quality Phrases from
  Massive Text Corpora}.
\newblock Special Interest Group on Management of Data, 2015.

\bibitem{D:6}
S.~Mazumder, B.~Bishnoi, and D.~Patel, {\em News Headlines: What They Can Tell
  Us?}
\newblock IBM Collaborative Academia Research Exchange, 2014.

\end{thebibliography}
\bibliographystyle{ieeetr}
\end{document}